\documentclass[aps,pre,twocolumn,groupedaddress,showpacs,floatfix]{revtex4}
\usepackage{graphicx}
\usepackage{amsmath}
\begin{document}
\title{Modeling friction on a mesoscale: Master equation for the
  earthquake-like model} 
\author{O.M. Braun}
\email[]{obraun@iop.kiev.ua}
\homepage[]{http://www.iop.kiev.ua/~obraun}
\affiliation{Institute of Physics, National Academy of Sciences of Ukraine,
  46 Science Avenue, 03028 Kiev, Ukraine}
\author{M. Peyrard}
\email[]{Michel.Peyrard@ens-lyon.fr}
\affiliation{Ecole Normale Sup\'{e}rieure de Lyon,
  46 All\'{e}e d'Italie, 69364 Lyon C\'{e}dex 07, France}
\date{\today}
\begin{abstract}
The earthquake-like model
with a continuous distribution of static thresholds
is used to describe the properties of solid friction.
The evolution of the model is reduced to a master equation which can be
solved analytically.
This approach naturally describes stick-slip and smooth sliding
regimes of tribological systems within a framework which separates the
calculation of the friction force from the studies of the properties
of the contacts.
\end{abstract}
\maketitle

In spite of its crucial practical importance, friction is still not
fully understood \cite{P0}. It raises questions at many scales, from
the atomic scale studied nowadays by atomic force microscopy
to the macroscopic scale of a solid block sliding on an
other. A simple mesoscopic model has been introduced to bridge the gap
in scales and describe the main experimental observations, such as
stick slip or smooth sliding, in terms of the properties of local
contacts. This widely used Burridge-Knopoff spring-block model \cite{BK1967},
initially introduced to study earthquakes (EQ model), has been
developed by Olami, Feder and Christensen \cite{OFC1992}. It describes
the contacts in terms of elastic springs and junctions that break at a
critical force. Computer simulations \cite{P1995,BR2002} showed that
the EQ model may 
reproduce the experimentally the observed stick-slip and
smooth-sliding regimes, including the role of velocity and
temperature, if the model is at least two-dimensional and various
assumptions on the properties of the contacts are made.

The drawback of such a simulation approach is that heavy calculations
with different parameter sets or contact properties are required to
determine the main features of the model, and it is hard to draw
conclusions of general validity. The calculations may be tedious
because a large number of contacts and investigations on very long
evolution times are necessary to get meaningful statistics and to make
sure that the calculation has reached asymptotic properties which are
not influenced by the initial conditions. Moreover almost all the studies
based on the EQ model assume for simplicity, and to reduce the
parameter space to explore, that all contacts have
identical properties. It turns out that, as we show below, this limit
is singular and may lead to qualitatively incorrect conclusions.
\begin{figure}[h] 
\includegraphics[clip,width=6cm]{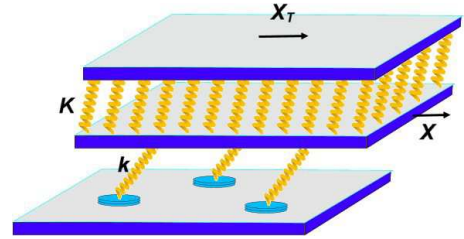} 
\caption{\label{A01}(color online): The earthquake-like model of friction.}
\end{figure}

\smallskip
Here we introduce a master equation (ME) approach which is much more
efficient than simulations and can be solved analytically in cases
which are particularly relevant. It provides a deeper understanding of
friction analyzed at the mesoscale in terms of the statistical
properties of the contacts. This splits the study of friction in two
independent parts: (i) the calculation of the friction force given by the
master equation provided the statistical properties of the contacts
are known, (ii) the study of the contacts and their statistics, which
needs inputs from the microscopic scale. Many aspects such as the
interaction between the contacts and their aging can be studied
separately to determine their role on the statistical properties of
the contacts and then accounted for by the master equation
approach.

\smallskip
\textit{Earthquake-like model}.
The EQ model describes the contact interface, i.e.\ the interface between
the bottom of the solid block and the fixed substrate 
(Fig.~\ref{A01}). It assumes that
the interaction occurs through $N_c$ asperities that make contacts
with the substrate. Each asperity is
characterized by its contact area $A_i$ and an elastic constant $k_i$,
schematized by an elastic spring on Fig.~\ref{A01}, which
can be estimated from $k_i \sim \rho \, c^2 \sqrt{A_i}$, where $\rho$
is the mass density and 
$c$ is the transverse sound velocity of the material which forms the
asperity \cite{P1995}. 
When the bottom of the solid block is moved by
$X$, the stretching $x_i$ of an asperity, i.e.\ its elastic deformation
with respect to its relaxed shape, increases. The force at the
contact grows as $f_i = k_i x_i$
until it reaches the threshold value $f_{si} \propto A_i$
at $x_{si} = f_{si} /k_i \propto \sqrt{A_i}$;
at this point the contact rapidly slides, and $f_i$ and $x_i$ drop to 
a small value before a contact is formed again.

Let $P_c (x_s)$ be the normalized probability distribution of values
of the thresholds $x_{si}$ at which contacts break. 
The model is studied in
the quasi-static limit where inertia effects are neglected so that only
positive values of $x_i$ are relevant.
The distribution $P_c (x)$ can be characterized by its average value
$\bar{x}_s$ and standard deviation $\sigma_s$.
A typical example is a Gaussian distribution $P_c (x) = G (x;
\bar{x},\sigma)$ \cite{note1}.

\begin{figure*}[t] 
\includegraphics[width=8.7cm, height=4cm]{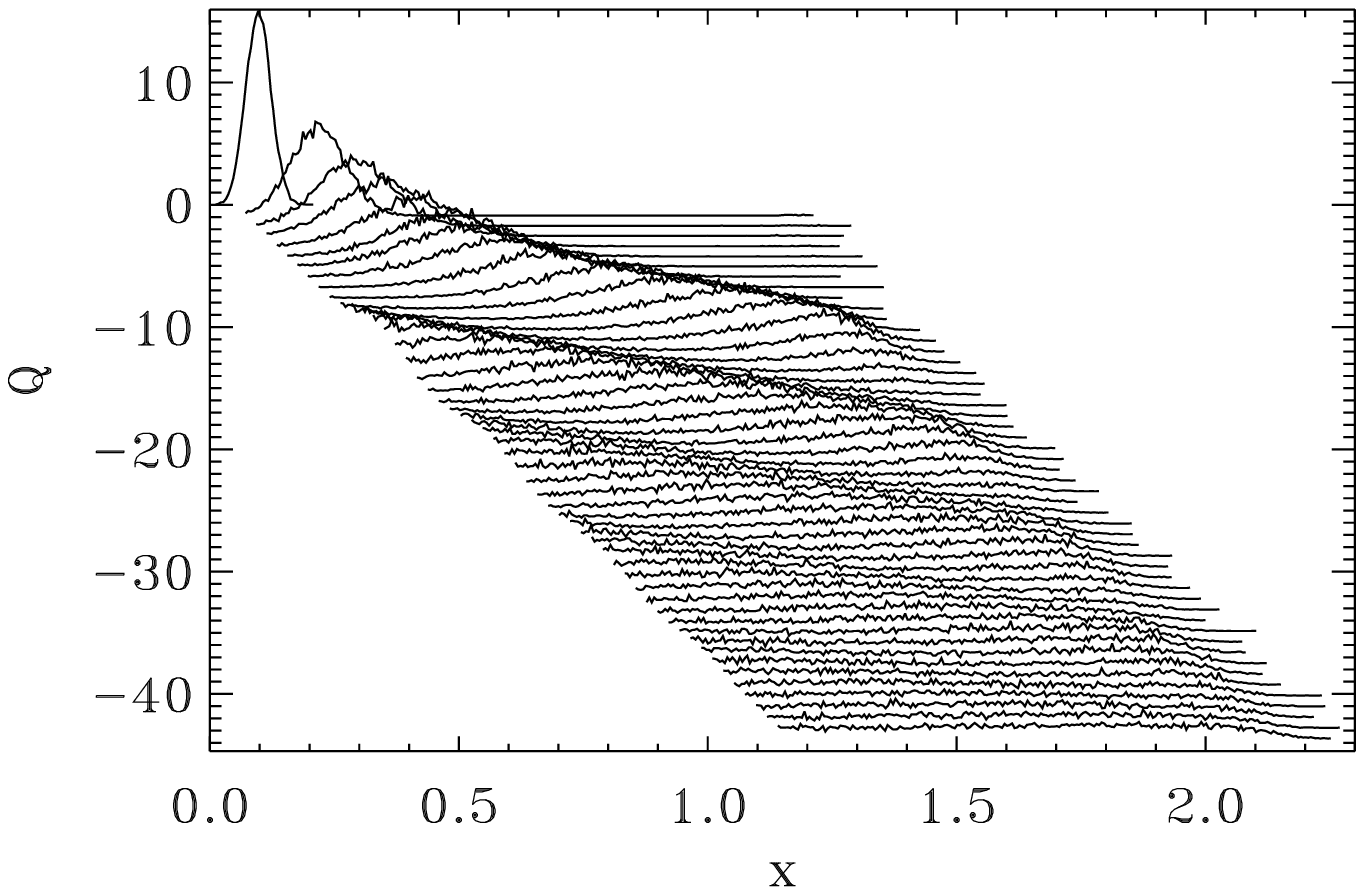} 
\includegraphics[width=8.7cm, height=4cm]{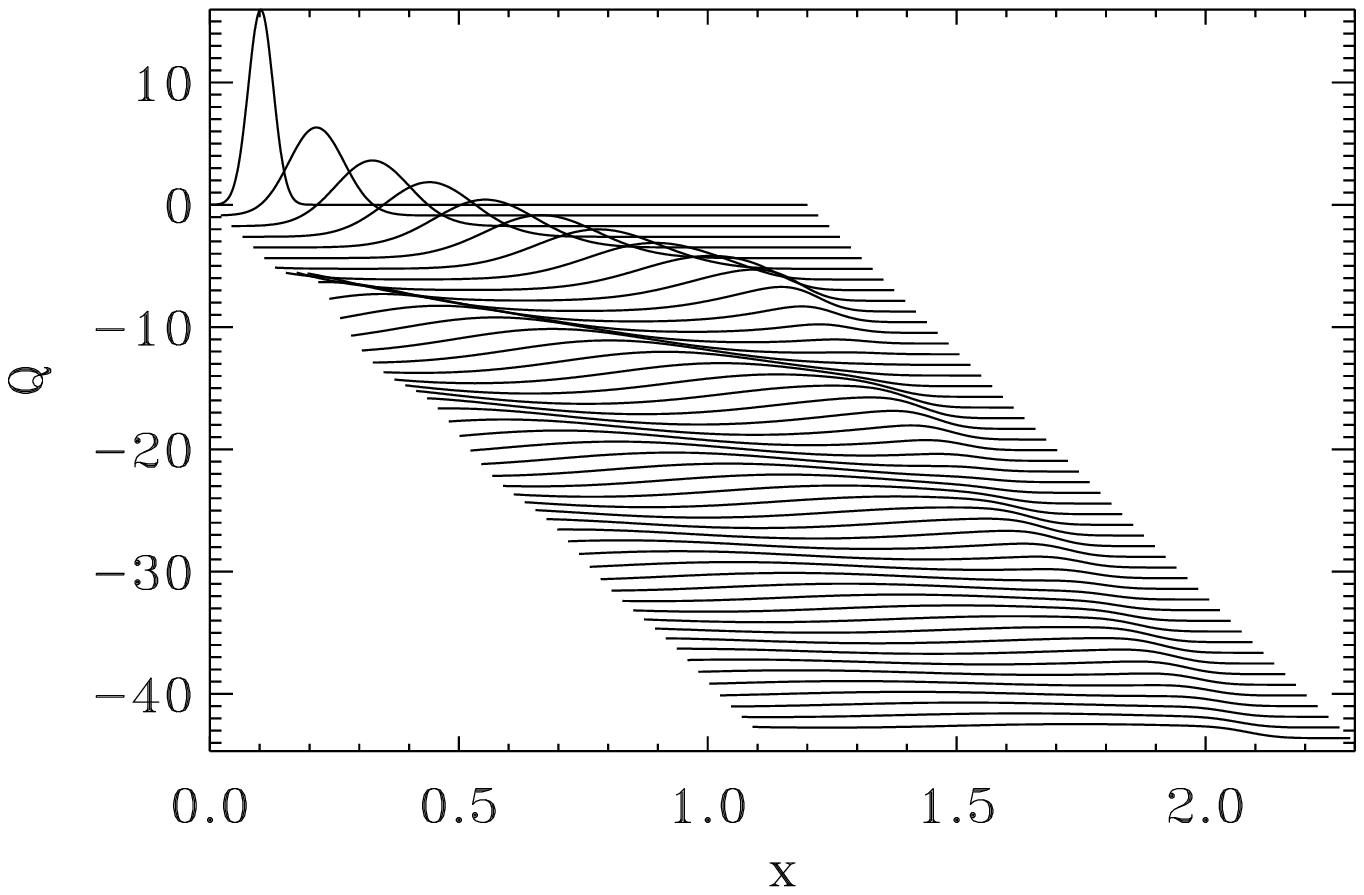} 
\includegraphics[width=7.6cm]{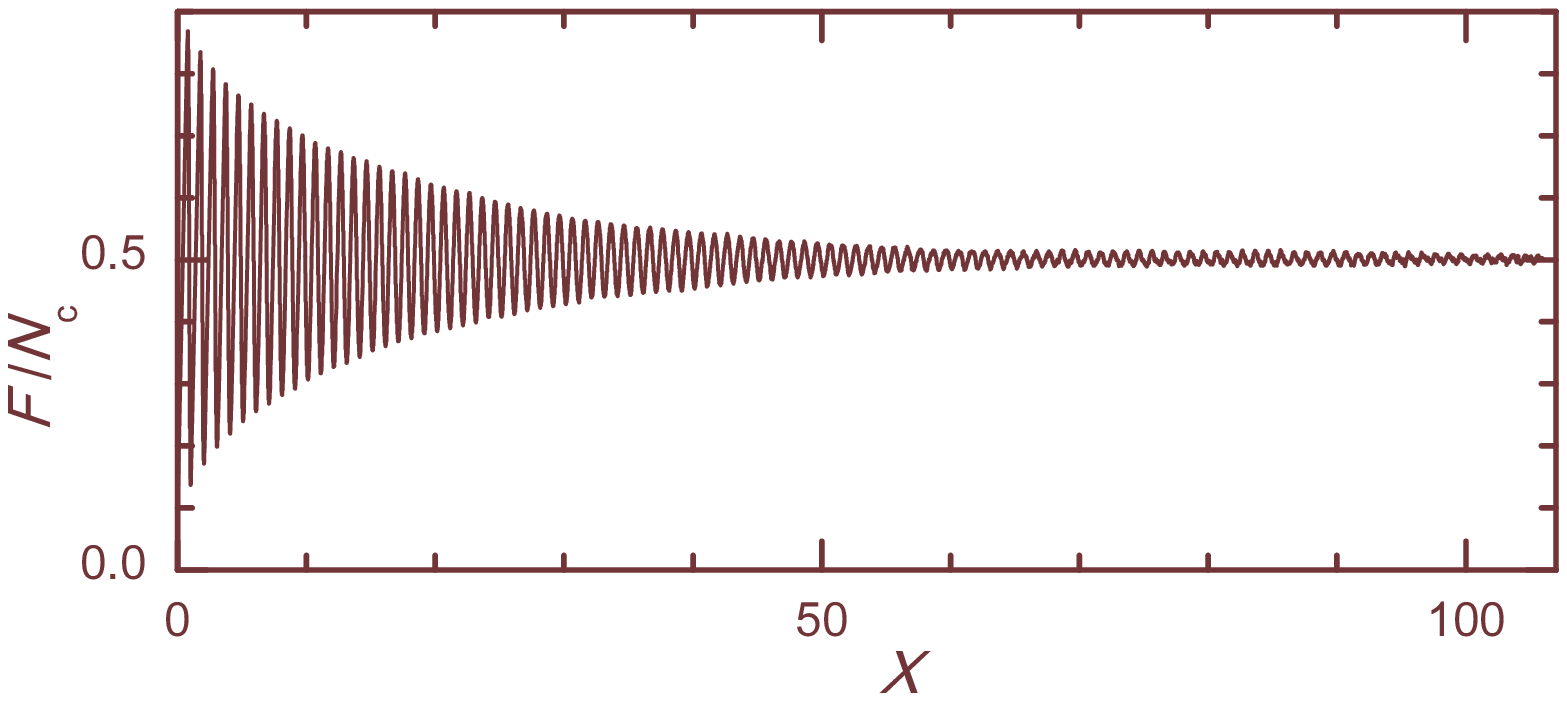}
\hspace{1.0cm}
\includegraphics[width=7.6cm]{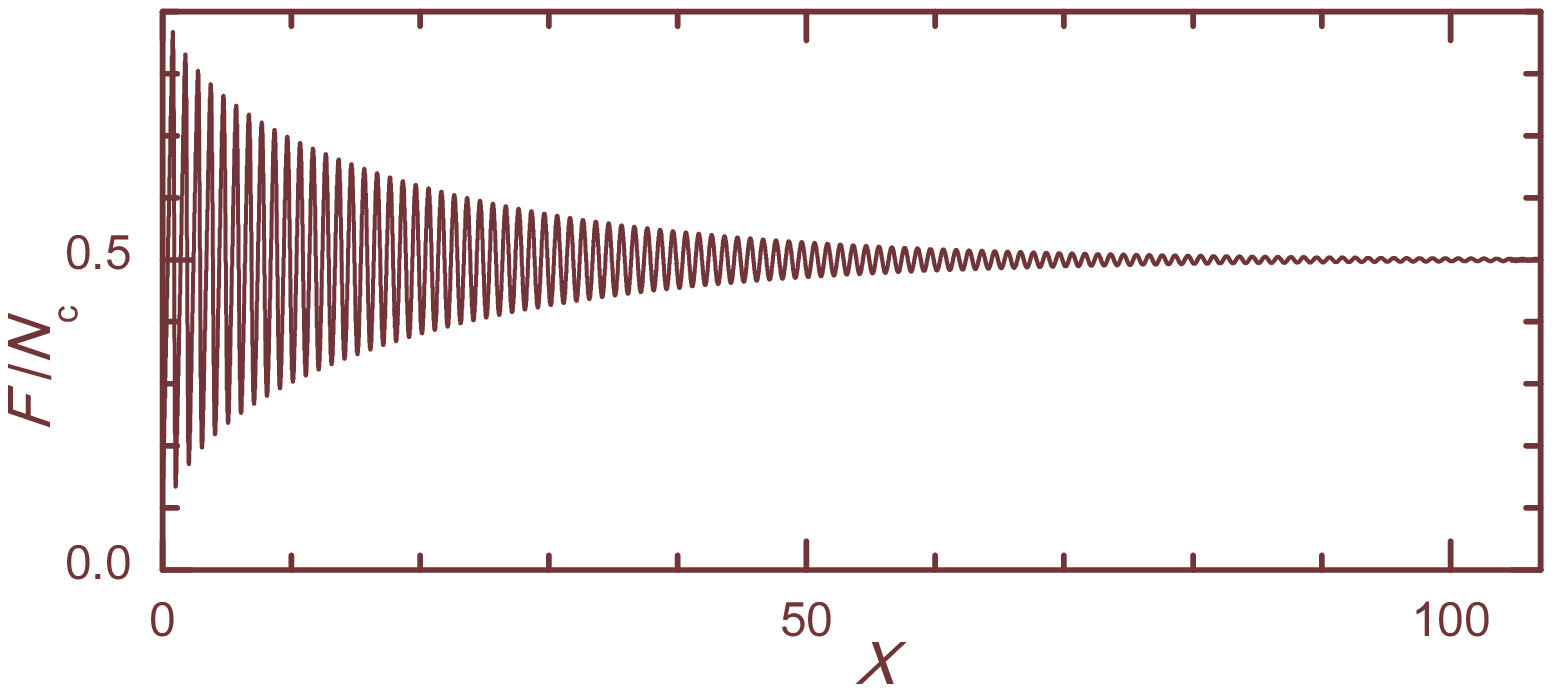}
\caption{\label{A02}
(a) Evolution of the EQ model.
The curves show the distribution $Q(x;X)$ for incrementally increasing
values of $X$ 
(with the step $\Delta X \approx 1$). 
The distribution $P_c (x)$ is Gaussian with
$\bar{x}_s =1$ and $\sigma_s =0.05$,
the initial distribution $Q_{\rm ini} (x)$ is
Gaussian with $\bar{x}_{\rm ini} =0.1$ and $\sigma_{\rm ini} =0.025$.
(b) Solution of the master equation with the increment $\Delta X =1.09$ for
the same model parameters. 
The bottom panels 
show the corresponding dependences $F(X)$ for $\langle k_i \rangle =1$ and
$K=\infty$.} 
\end{figure*}

To describe the evolution of the model,
we introduce the distribution $Q(x;X)$ of the stretchings $x_i$ 
when the bottom of the solid block is at a position $X$. It is normalized 
by $\int_{0}^{\infty} dx \, Q(x;X) = 1$ for all $X$.
Let all asperities be initially relaxed or weakly stressed, 
e.g., let the distribution $Q(x;0)=Q_{\rm ini} (x)$
be the Gaussian $Q_{\rm ini} (x) = G(x; \bar{x}_{\rm
  ini},\sigma_{\rm ini})$ 
with $\bar{x}_{\rm ini} \ll \bar{x}_s$ and $\sigma_{\rm ini} \alt \sigma_s$.
Now, let us adiabatically increase the displacement $X$ of the bottom of the
solid block while the substrate remains fixed.
The sum of the elastic forces exerted on the bottom of the block
by the stretched asperities makes up the {\em friction force}
\begin{equation}
F(X)=N_c \langle k_i \rangle \int_{0}^{\infty}  x \, Q(x;X) dx \; .
\label{Q6}
\end{equation}

The evolution of the system, deduced from the numerical simulation of
the EQ model is shown in Fig.~\ref{A02}a.
It shows that, in the long term, 
the initial distribution approaches a
stationary distribution $Q_s (x)$ and
the total force $F$ becomes independent on $X$. The final
distribution is independent of the initial one.
An elegant mathematical proof of this statement was presented in
Ref.~\cite{FDW2005}.
The statement is valid for any distribution $P_c (x)$
except for the singular case of $P_c (x) = \delta (x-x_s)$.

\smallskip
\textit{Master equation}.
Rather than studying the evolution of the distribution $Q(x;X)$ by a
simulation of the EQ model it is possible to describe it {\em analytically}.
Let us consider a small displacement $\Delta \!
X$ of the bottom of the solid block. It induces a variation of the
stretching $x_i$ of the asperities which has the same value  $\Delta
\! X$ for all asperities if the deformation of the bottom surface of the
block can be neglected. As discussed below, the general case where the
relative positions 
of the asperities on the surface are allowed to vary can be cast into
a generalization of this formalism.
The displacement $X$ leads to three kinds of changes in the
distribution $Q(x;X)$: first, there is a shift due to the global
increase of the stretching of the asperities, 
second, some contacts break because the stretching exceeds the
maximum that they can stand,
and third, those broken contacts form again,
at a lower stretching, after a slip at the scale of the asperities,
which locally reduces the tension within the corresponding
asperities.
These three contributions can be written as a master equation for $Q(x;X)$:
\begin{equation}
Q(x;X + \Delta \! X) = Q(x - \Delta \! X;X) - \Delta Q_-(x;X) + \Delta
Q_+(x;X). 
\label{Q2}
\end{equation}
The first term in the r.h.s.\ of Eq.~(\ref{Q2}) is just the shift.
The second term $\Delta Q_-(x;X)$ designates the variation of the
distribution due to the 
breaking of some contacts. It can be written as
\begin{equation}
\Delta Q_-(x;X) = P(x) \, \Delta \! X \, Q(x;X)  \; ,
\label{Q3}
\end{equation}
where $P(x) \, \Delta \! X$ is the fraction of contacts that break
when the position changes from
$X$ to $X + \Delta \! X$. According to the definition of $P_c(x)$ the
total number of unbroken contacts when the stretching of the
asperities is equal to $x$ is given by $N_c \int_x^\infty P_c(\xi) \; d\xi$. 
The contacts that break when $X$ increases by $\Delta X$,
so that the stretching of all asperities increases by $\Delta X$, is the
number of contacts which have their thresholds between $x$ and $x +
\Delta X$, i.e.\ $N_c P_c(x) \Delta X$. Thus
\begin{equation}
P(x) = P_c (x) \Big/ \int_{x}^{\infty} d\xi P_c (\xi) \; .
\label{PcP}
\end{equation}
The broken contacts relax and have to be added to the distribution
around $x \sim 0$, leading to the third term in Eq.~(\ref{Q2}).
We denote by $R(x)$ the normalized distribution of stretchings 
for the relaxed contacts.
Writing that all broken contacts described by $\Delta Q_-(x;X)$
reappear with the distribution $R(x)$, we get
\begin{equation}
\Delta Q_+(x;X) = R(x) \int_{0}^{\infty} d\xi \, \Delta Q_-(\xi;X).
\label{Q4}
\end{equation}
Equation (\ref{Q2}) can be rewritten as
$[ Q(x;X + \Delta \! X) - Q(x;X) ] + [ Q(x;X) - Q(x - \Delta \! X;X) ]
 =  - \Delta Q_-(x;X) + \Delta Q_+(x;X)$.
Taking the limit $\Delta \! X \to 0$, we finally get the
integro-differential equation
\begin{eqnarray}
\frac{ \partial Q(x;X) }{ \partial x} +
\frac{ \partial Q(x;X) }{ \partial X} +
P(x) \, Q(x;X)
\nonumber \\
= R(x) \int_{0}^{\infty} d\xi \, P(\xi) \, Q(\xi;X),
\label{Q5a}
\end{eqnarray}
which has to be solved with the initial condition
$Q(x;0) = Q_{\rm ini} (x)$.
Notice that  $Q_{\rm ini} (x)$ cannot be an arbitrary function,
because the contacts that exceed their stability threshold, must relax
from the very beginning. 

Once the distribution $Q(x;X)$ is known, we can calculate the 
friction force $F(X)$
from Eq.~(\ref{Q6}).
The static friction force is the maximum of $F(X)$, i.e.,
$F_s = F(X_s)$, where $X_s$ is a solution of the equation
$F^{\prime}(X) \equiv dF(X)/dX =0$.
In order to simplify the calculation,
we will assume in what follows that $R(x)=\delta(x)$, i.e. when the broken
contacts stick again the asperities are completely relaxed.

\begin{figure} 
\includegraphics[clip, width=8cm]{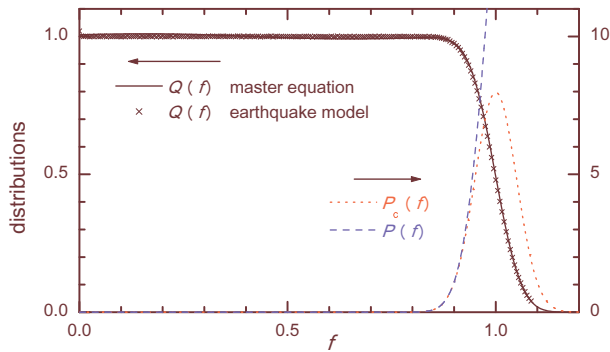}
\caption{\label{A03c}(color online):
The final distribution $Q(x)$ for the parameters from Fig.~\ref{A02}
(solid curve; crosses show the averaged final distribution for the EQ model).
The red dotted curve shows the distribution $P_c (x)$, and the blue broken
curve shows $P(x)$.} 
\end{figure}

\smallskip
Analytical solutions of the master equation can be obtained for some
particular cases of contact properties \cite{BPnext}, such as 
a rectangular $P_c (x)$ distribution.
Moreover, for one particular but important choice of the initial distribution,
when all contacts are relaxed at the beginning, $Q_{\rm ini} (x) = \delta (x)$,
we can find analytically the initial part of the solution in a general 
case \cite{BPnext}.
For the rather general case of a Gaussian distribution of thresholds,
$P_c (x) = G(x; \bar{x}_s, \sigma_s)$,
a numerical solution of the master equation (\ref{Q5a}) is presented in
Fig.~\ref{A02}b. 
One can see that it is almost identical to that of the EQ model
(Fig.~\ref{A02}a), 
except for the noise on the EQ distributions.
The distribution $Q(x;X)$ always approaches a stationary
distribution $Q_s (x)$. 
The final distributions of the EQ model and the master equation approach
are compared in Fig.~\ref{A03c}.

\smallskip
The steady-state, or smooth-sliding solution, i.e.\ the solution 
of Eq.~(\ref{Q5a}) which
does not depend on $X$, 
can easily be found \cite{P1995}.
In a general case it can be expressed as 
\begin{equation}
Q_s (x) = C^{-1} \Theta(x) E_P (x),
\label{Q11a}
\end{equation}
where $\Theta(x)$ is the Heaviside step function, 
$E_P (x) = e^{-U(x)}$, $U(x) = \int_{0}^{x} d\xi \, P(\xi)$, and
$C = \int_{0}^{\infty} dx \, E_P (x)$.
The (kinetic) friction force is then equal to
\begin{equation}
F = (N_c /C) \langle k_i \rangle \int_{0}^{\infty} dx \, x E_P (x).
\label{Q11d}
\end{equation}

\smallskip
In the general case, let the distribution $P_c (x)$ be of bell-like shape
with the maximum at $\bar{x}_s$ and the width $\sigma_s$.
When $X$ shifts for the distance $\bar{x}_s$, due to the breaking and
reforming of contacts with a lower stretching,
an initially peaked distribution $Q(x;X)$ broadens by the value $\sim
\sigma_s$ (Fig.~\ref{A02}).
Therefore, any initial distribution tends to the stationary one as
$| Q(x;X) - Q_s (x) | \propto \exp (-X/X^*)$, where $X^* \sim
\bar{x}_s^2 / \sigma_s$. 

Thus, in a general case the solution of the master equation 
always approaches the
smooth-sliding one given by Eq.~(\ref{Q11a}). 
However, there is one exception from this general scenario.
When all contacts are identical, i.e., all contacts are characterized by
the same threshold $x_s$, $P_c(x) = \delta (x-x_s)$, the model admits
a periodic solution \cite{BPnext}. 
This {\em singular} periodic solution has been found in simulations
and analyzed as describing the
stick slip \cite{BN2006}, 
but actually it is unphysical and ceases to exist as
soon as non-equivalent contacts are considered, whatever their precise
properties. As discussed below the stick slip can be deduced from the
solution of the master equation, but its origin is different
\cite{Caroli}. 

\bigskip
\textit{Stick slip and smooth sliding.} The
master equation allows us to compute the friction force $F(X)$ when
the bottom of the solid block is displaced by $X$. But actually we don't
control $X$. The
displacement is caused by a shearing force $F_T$ applied on the top of
the solid block which displaces the top surface by $X_T$. As the strain
on the solid is generally small the deformation of the solid block can
be assumed to be elastic so that $X_T$ is related to the applied force
by $F_T = K (X_T - X)$ where $K$ is the shear elastic constant of
the solid block. The total force
applied to the bottom of the solid block, which determines its
displacement $X$, is the 
sum of the applied force and the friction force
\begin{equation}
\label{eq:totforce}
 F_{\rm tot} = K (X_T - X) - F(X) \; .
\end{equation}
It can be viewed as deriving from the potential
\begin{equation}
\label{eq:pottot}
V(X_T,X) = \frac{1}{2} K \, (X_T - X)^2 + \int_0^X F(\xi) d\xi
\; ,
\end{equation}
which determines the behavior of the solid block subjected to friction
and applied force. A necessary condition for smooth sliding is that
$X_T$ and $X$ grow together with $X_T - X = B$, where $B$ is a
constant that measures the shear strain of the solid block during the
sliding. It is determined by the condition
  ${\partial V}/{\partial (X_T - X)} = - {\partial
  V}/{\partial X} = 0\;$ ,
which simply means that the total force on the interface vanishes. Smooth
sliding also requires this state be stable, 
\begin{equation}
\label{eq:stab}
\dfrac{\partial^2 V}{\partial (X_T - X)^2} = 
\dfrac{\partial^2 V}{\partial X^2} \ge 0 \quad {\mathrm{or}} \quad F'(X)
\ge - K \; .
\end{equation}
If we start from relaxed asperities, in the early stage of the motion
$F(X)$ is a growing function of $X$, and then it passes by a maximum
when some contacts start to break and reform at lower asperity
stress. As a result $F'(X)$ becomes negative. Depending on the value of $K$
two situations are possible. For large $K$ (stiff block) $F'(X)$ never
falls below $- K$ and the smooth sliding is a stable steady
state. For small $K$ (soft block) $F'(X)$ can become smaller than $-
K$ so that the stability condition (\ref{eq:stab}) is no longer
valid. The instability causes $X_T - X$ to change abruptly by a
breaking of all the contacts and a quick slip of the block before the
contacts reform with relaxed asperities. And the process can repeat
again, leading to the familiar stick slip motion. The master
equation, which gives $F(X)$ can be used to compute the period of the
stick slip, and, when the asperities fully
relax before the contacts reform, an analytical
solution can be obtained \cite{BPnext}. It should be noticed that the
existence of a stick slip is not only determined by the stiffness $K$
of the solid block. The properties of the asperities, defined by the
distribution $P_c(x)$ of the stretching for which they break  is also
essential because it determines the expression of $F(X)$ and hence
the minimal value of $F'(X)$.

\medskip

\textit{Discussion.}
The ME formalism can be extended to take into account various
generalizations of the EQ model. For instance, in establishing the
master equation we assumed that the asperities were fixed with respect
to the solid block and were moving together with it. This is an
approximation and one can, in principle, take into account the elastic
deformation of the interface by introducing for instance
a {\em position dependent}
distribution $Q[x,X(\vec{r})]$ where $\vec{r}$ denotes
the position of an asperity on the interface, and $X(\vec{r})$ the
local translation of the interface averaged over a mesoscopic scale.
The master equation
must then be coupled to an equation describing the elastic deformation
of the interface, subjected to the contact forces at each point
$\vec{r}$. This illustrates the new viewpoint introduced
by the ME approach which describes the phenomena at an intermediate
scale between the microscopic scales of the contacts and the
macroscopic scale of the displacement of the solid block. The ME
equation only has a meaning if, on this intermediate scale, there are
many individual contacts, allowing us to study them as a statistical
distribution and not individually. A simpler view of the effect of the
elastic interaction between the contacts is to describe it as a
renormalization of the distribution $P_c(x)$ in a
mean field like approach.

Another aspect which can be introduced in the ME formalism is the
aging of the contacts \cite{BPnext}. Experiments and MD simulations
show that the static friction force grows with time since a contact is
formed. As a result, if newly formed contacts are characterized by a
distribution $P_{ci}(x)$ this distribution evolves with time,
moving to a higher mean value and a smaller standard deviation. 
Aging is a stochastic process which can be described by a Smoluchowsky
equation.
The master equation for $Q(x,X)$ must then be
completed by an equation for the evolution of $P_c(x)$, which in turns
affects
$P(x)$ in the master equation. Time enters through the velocity of the
sliding, a faster sliding giving less time for the aging of the
contacts. The full development is too long to be given
here \cite{BPnext} but it is easy to realize that a larger velocity
leads to a smaller friction force because the contacts have less time
to age, which is the source of a potential instability. Aging is
therefore another cause for the stick-slip behavior.

The ME formalism can also accommodate temperature effects which enter
again through their effect on the distribution $P(x)$ because thermal
fluctuations allow an activated breaking of contacts for asperities
which are still below their thresholds \cite{P1995}. The main point that 
we would like to stress is
that this formalism introduces a new viewpoint on the mescoscopic
modeling of friction. Moreover it splits the analysis of friction
phenomena into problems that can be studied separately, the
statistical properties of the contacts, and the evolution of the
distribution $Q$ which is described by the master equation,
coupled to additional equations representing different effects such
as the elastic interactions between asperities, the aging of the
contacts or temperature fluctuations.

\smallskip
This work was supported by CNRS-Ukraine grant No.~18977.
O.B.\ acknowledges a partial support from the EU Exchange Grant
within the ESF program ``Nanotribology'' (NANOTRIBO).



\begin{thebibliography}{xx}
\bibitem{note1} 
More rigorously, the distribution $P_c (x)$ should be truncated to
be defined on the interval
$0 < x < x_m$, where $x_m \gg \bar{x}_s$
\end{thebibliography}
\end{document}